\documentclass[3p,times,review]{elsarticle} 

\usepackage[dvipsnames,table]{xcolor}

\usepackage{amsmath}
\usepackage{siunitx}

\usepackage[ruled,linesnumbered]{algorithm2e}

\usepackage{booktabs}
\usepackage{subcaption}
\usepackage{url}
\usepackage[hidelinks]{hyperref}

\usepackage{array,makecell}

\usepackage[inline]{enumitem}

\usepackage[capitalise,nameinlink,sort]{cleveref}

\usepackage{amssymb}

\journal{Additive Manufacturing}

\begin{document}

\begin{frontmatter}



\title{Force Controlled Printing for Material Extrusion Additive Manufacturing}


\author[inst1,inst2]{Xavier Guidetti}
\ead{xaguidetti@control.ee.ethz.ch}
\author[inst4]{Nathan Mingard}
\ead {nathan.mingard@students.fhnw.ch}
\author[inst1]{Raul Cruz-Oliver}
\ead{rcruzoliver@student.ethz.ch}
\author[inst3]{Yannick Nagel}
\ead{yannick.nagel@nematx.com}
\author[inst2]{Marvin Rueppel}
\ead {marvin.rueppel@inspire.ch}
\author[inst5]{Alisa Rupenyan}
\ead {alisa.rupenyan@zhaw.ch}
\author[inst1,inst2]{Efe C. Balta\corref{cor1}}
\ead{efe.balta@inspire.ch}
\author[inst1]{John Lygeros}
\ead{lygeros@control.ee.ethz.ch}

\affiliation[inst1]{organization={Automatic Control Laboratory, ETH Zurich},
            addressline={Physikstrasse 3}, 
            postcode={8092}, 
            city={Zurich},
            country={Switzerland}}
            
\affiliation[inst2]{organization={Control and Automation Group, Inspire AG},
            addressline={Technoparkstrasse 1}, 
            postcode={8005}, 
            city={Zurich},
            country={Switzerland}}
            
\affiliation[inst3]{organization={NematX AG},
            addressline={Vladimir-Prelog-Weg 5}, 
            postcode={8093}, 
            city={Zurich},
            country={Switzerland}}

\affiliation[inst4]{organization={Institute of Polymer Engineering, FHNW},
            addressline={Klosterzelgstrasse 2}, 
            postcode={5210}, 
            city={Windisch},
            country={Switzerland}}

\affiliation[inst5]{organization={ZHAW Centre for AI, Zurich University of Applied Sciences},
            addressline={Technikumstrasse 71}, 
            postcode={8400}, 
            city={Winterthur},
            country={Switzerland}}

\cortext[cor1]{Corresponding author}

\begin{abstract}

In material extrusion additive manufacturing, the extrusion process is commonly controlled in a feed-forward fashion. The amount of material to be extruded at each printing location is pre-computed by a planning software. This approach is inherently unable to adapt the extrusion to external and unexpected disturbances, and the quality of the results strongly depends on a number of modeling and tuning parameters. To overcome these limitations, we propose the first framework for Force Controlled Printing for material extrusion additive manufacturing. We utilize a custom-built extruder to measure the extrusion force in real time, and use this quantity as feedback to continuously control the material flow in closed-loop. We demonstrate the existence of a strong correlation between extrusion force and line width, which we exploit to deposit lines of desired width in a width range of \SI{33}{\percent} up to \SI{233}{\percent} of the nozzle diameter. We also show how Force Controlled Printing outperforms conventional feed-forward extrusion in print quality and disturbance rejection, while requiring little tuning and automatically adapting to changes in the hardware settings. With no adaptation, Force Controlled Printing can deposit lines of desired width under severe disturbances in bed leveling, such as at layer heights ranging between \SI{20}{\percent} and \SI{200}{\percent} of the nominal height.

\end{abstract}




\end{frontmatter}


\section{Introduction} \label{sec:intro}

Additive Manufacturing (AM), commonly known as 3D printing, is a prominent technique that enables manufacturing of 3D objects with complex geometrical features by depositing material layer by layer. Material extrusion additive manufacturing \cite{ASTM_standard}, better known as Fused Filament Fabrication (FFF), or Fused Deposition Modeling (FDM), is one of the most widespread and popular AM techniques \cite{sculpteo2021} due to its accessibility and versatility. It is used in rapid prototyping and small-lot manufacturing in a variety of sectors and is successful among professional users as well as hobbyists. In FFF, a feedstock thermoplastic material (generally in the form of a filament) is pushed by a filament-driving wheel into a heated nozzle, from which the material exits as a melted plastic bead. This assembly called an extruder, is moved in space by a set of numerically controlled motion drives to deposit plastic at desired locations and form the final part \cite{gibson2021additive}. A software known as \emph{slicer} is used to produce commands for the printer. Based on the desired geometry and various settings, the software computes the set of trajectories required to produce a part. One central challenge lies in designing the commands sent to the extruder motor, which have to be synchronized with the machine motion to take into consideration the dynamics of the extrusion process to achieve high-quality results \cite{shaqour2021gaining}. 

Numerous works have developed models of the extrusion process dynamics \cite{moretti2021process,bellini2004liquefier,turner2014review}, and used them to optimize the feedforward input sent to the machine and improve performance \cite{comminal2019motion,wu2023modeling,habbalsystem2024}. Particularly when targeting high geometrical accuracy, it was shown that performance can be improved with accurate geometric modeling of the deposited material beads as a function of the print parameters \cite{balta2022numerical,aksoy2020control}. All these approaches, however, suffer from the classical limitations of feed-forward planning \cite{broussard1980feedforward}. The pre-computed commands have no adaptability to changes in the printer dynamics or material properties, are highly susceptible to external and unexpected disturbances, and strongly depend on calibration and modeling. The complex and non-linear nature of the extrusion process especially amplifies the modeling-related issues. 

Despite the evident shortcomings of feed-forward printer control, this technique remains state-of-the-art in FFF due to the complexities of collecting real-time process measurements for closed-loop control. To this end, numerous recent works explore the field of in-situ monitoring for FFF at different time-scales~\cite{rao2015online}. In-situ monitoring has been demonstrated and applied in 3D printing health management~\cite{kim2018development}, quality monitoring~\cite{chhetri2019quilt}, clog detection~\cite{tlegenov2017dynamic}, process monitoring \cite{anderegg2019situ}, and anomaly detection~\cite{balta2019digital,li2019situ}. Despite the recent developments in process monitoring in practice and literature, online control of FFF is lagging. Applications have appeared particularly at slower control time scales. For example, measurements conducted in between the deposition of two layers have been used for layer-to-layer accurate process modeling~\cite{balta2021layer} and optimal parameter adaptation~\cite{GuidettiIFAC}. Layer-to-layer approaches collect data and apply correction only at a low sampling rate between the layers. As a result, while these approaches are a step forward from purely open-loop feed-forward control, they remain inefficient against fast-acting run-time disturbances in the extrusion process during the deposition of a layer.  

When designing a fast and accurate printing process, counteracting disruptions in the material flow and deposition is required to produce high-quality parts \cite{serdeczny2018experimental,serdeczny2019numerical,serdeczny2020numerical}. Recent studies have begun investigating the possibility of collecting high-frequency measurements about the extrusion process itself, such as the material flow in the nozzle \cite{coogan2019line}, the force measured in the extruder while printing in the air \cite{fischer2023line}, or the filament feed \cite{moretti2020towards}. Furthermore, some works have conducted closed-loop control of the extrusion process based on run-time data measured while printing \cite{moretti2023closed, greeff2017closed}. These studies have focused on avoiding slippage between the driving wheel and the filament, and while this represents an improvement in process control, the measurements utilized are not able to fully characterize the extrusion process. Specifically, by measuring the filament feed, it is not easily possible to characterize the disturbances that the interactions between the filament and the nozzle or print bed may cause.

In this work, we propose the first framework for Force Controlled Printing (FCP) for FFF. We develop a custom-built and highly sensitive sensor configuration for FFF and utilize it to collect run-time information that characterizes the extrusion process and run-time disturbances. We abandon the existing open-loop approach to extrusion planning in favor of closed-loop control based on the force required to extrude material. Technical aspects related to controller tuning are available in our preliminary study \cite{guidetti2024print}. 
We use a high-performance extrusion controller to demonstrate and study the advantages of FCP over conventional FFF. The main contributions of this work are:
\begin{itemize}[leftmargin=*]
    \item The introduction of a novel FCP hardware setup and framework for FFF;
    \item The illustration of a disturbance-free method to print lines of desired width over a wide width range using FCP;
    \item The experimental demonstration that FCP outperforms conventional FFF in print quality and disturbance rejection across practical use cases such as inconsistent layer heights within a layer, extruder drive slippages, and surface defects.
\end{itemize}
The paper structure is as follows: Sec. \ref{sec:matmet} introduces the material utilized and details the sensing and control configuration for FCP; in Sec. \ref{sec:calc} we tune and validate the FCP controller; finally, Sec. \ref{sec:results} and \ref{sec:disc} showcase the obtained results and discuss them.

\section{Material and methods} \label{sec:matmet}

\subsection{Material}

In this study, we utilize as feedstock material Liquid Crystal Polimers (LCP). The possibility to utilize this material in FFF has been first introduced in \cite{gantenbein2018three}, and LCP filaments are currently commercialized by NematX AG\footnote{\url{https://nematx.ch}} under the name Nema HSS. They are used in high-performance applications requiring high precision or mechanical strength~\cite{guidetti2023stress}. LCPs are composed of aromatic thermotropic polyesters that self-assemble into nematic domains when heated above their melting temperature. In each nematic domain, the molecules have their long axes arranged in parallel. Extrusion through a thin heated nozzle was shown to produce global alignment of the molecules since the deformations produced during the extrusion align the nematic domains in the direction of extrusion. After cooling, the monomers are frozen in place and remain aligned in the axial direction of the deposited line. This confers extraordinary mechanical properties to LCP printed parts when mechanical stress is applied in the direction of filament deposition (see Fig. 3 of \cite{gantenbein2018three}). However, LCPs are sensitive to the parameters utilized during the extrusion process and during the filament deposition \cite{GuidettiIFAC}. Additionally, both over- and under-extrusion have been shown to strongly affect the mechanical properties of printed parts \cite{GuidettiIFAC,siqueira2017}. Thus, it is of crucial importance to control perfectly the extrusion process and the amount of deposited material to fully exploit the properties of LCPs and produce high-end components.

We remark that while our study primarily demonstrates applications of FFF with LCP material, the presented FCP framework is general purpose and can be used with any material choice with proper parameter tuning and process design.

\subsection{Methods}

\subsubsection{Extrusion Force} \label{sec:extr_force}

To feed filament into an extruder, a driving wheel in contact with the filament is actuated by a motor (see~\cref{fig:printer}). As the wheel rotates at a given speed, a feeding force $F_e$ is applied to the filament. This feeding force is counterbalanced by several opposing forces \cite{serdeczny2018experimental}. Simplifying the extrusion process and defining the filament displacement direction as the Z-axis, the following force equilibrium holds
\begin{equation}
    F_e = (F_t + F_p) + F_s = F_n + F_s \,, \label{eq:extr_force}
\end{equation}
where $F_t$ are forces due to tangential stresses on the inside of the nozzle in the Z-direction, $F_p$ are surface forces in the Z-direction produced by restrictions in the nozzle (such as the nozzle cone), and $F_s$ is the normal force produced by the print bed (or substrate) on the melted filament \cite{fischer2023line}. We regroup $F_t$ and $F_p$ in $F_n$ which represents the forces produced by the filament-nozzle interaction, in the Z-direction. Ultimately, the force $F_e$ required to move the filament through the heater and nozzle and to press the plastic bead on the print bed will depend on extrusion-related parameters (such as material feed rate, temperature, and filament and nozzle geometry, whose effect appears in $F_n$), and on substrate related parameters (such as layer height, print bed leveling, and substrate roughness, whose effect appears in $F_s$).

\subsubsection{Sensing Extruder} \label{sec:printhead}

\begin{figure}[htbp]
\centerline{\includegraphics[width=0.49\textwidth]{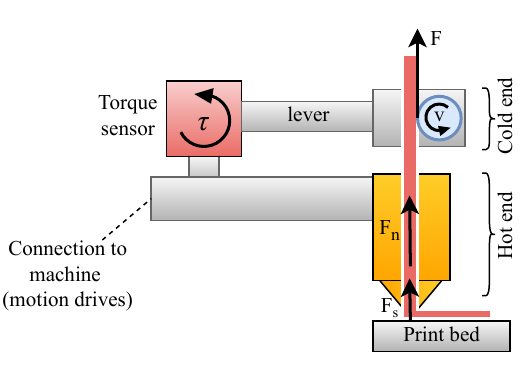}}
\caption{Schematic representation of the sensing extruder. The torque sensor is used to characterize the extrusion process.}
\label{fig:printhead}
\end{figure}

To obtain real-time measurements that characterize the extrusion process, we have developed an extruder assembly that includes a custom-built torque sensor. The printer head is represented schematically in~\cref{fig:printhead}. Contrarily to a conventional extruder, where the \emph{hot end} (heater and nozzle) and the \emph{cold end} (filament driving wheels and motor) are rigidly connected to each other, we utilize a configuration where the cold end is attached to the rest of the extruder assembly through a sensor. In particular, the driving wheels and motor are mounted on the free end of a lever, while the lever pivot is constituted by a torque sensor. During extrusion, as the filament driving wheel rotates, a force $F_e$ (see Eq. \eqref{eq:extr_force}) is applied onto the filament. This force is matched by a reaction force 
\begin{equation}
    F=F_e \label{eq:balance}
\end{equation}
of opposite direction which pushes the cold end away from the hot end. As shown in \cref{fig:printhead}, our extruder configuration enables one to compute the reaction force acting on the cold end by measuring the torque at the lever pivot point. For a torque $\tau$ and a lever of length $l$, it is trivial to retrieve the extrusion reaction force as $F = \tau/l$. The hot end is rigidly connected to the printer body, and consequently, the printing quality under nominal conditions is identical to the one achieved on a standard commercially available extruder. This is not the case with other existing setups. For example, in \cite{fischer2023line} the hot end is suspended to a force sensor and can thus move; in addition, the measurements are affected by the contact between the nozzle tip and previously deposited material. In \cite{anderegg2019situ}, a pressure sensor is installed in the nozzle to retrieve information about the extrusion process. This configuration is mechanically identical to a conventional extruder, but the sensor is cumbersome and its sensing interface (which creates sharp edges inside the nozzle) is known to interfere with the flow of technical polymers.

\subsubsection{Extrusion Control}

In the conventional open-loop approach to FFF, the filament flow rate is precomputed offline by software (generally known as slicer) that calculates the theoretically required amount of material as a function of various parameters, such as layer height, nozzle diameter, print speed, line width. The calculated flow rate (matched with a motion trajectory) is given to the machine as a sequence of instructions, which the printer follows to produce a part. The performance achievable by this approach is inherently limited by its open-loop nature, as discussed in \cref{sec:intro}. In this work, we propose a method for closed-loop control of extrusion, where the filament flow rate is computed online based on the reaction force produced while printing. 

\begin{figure}[htbp]
\centerline{\includegraphics[width=0.49\textwidth]{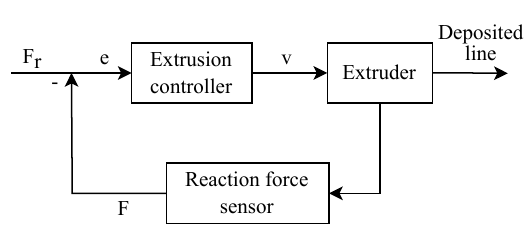}}
\caption{Block diagram of the closed-loop extrusion control scheme utilized in Force Controlled Printing}
\label{fig:control_block}
\end{figure}

\Cref{fig:control_block} shows the closed-loop control block diagram which we use to print at a desired reaction force. Feedback from the torque sensor is used to compute the difference between a reference reaction force and the measured reaction force. Based on this difference, the controller adapts the material flow rate. Intuitively, if the measured reaction force drops below the reference, a higher material flow rate is requested; conversely, when the measured reaction force is larger than the reference, the material flow rate is reduced. 

To achieve satisfactory performance, we use a proportional-integral-derivative (PID) controller \cite{visioli2006practical} that has been modified to include two derivative terms. One derivative term is applied directly to the error derivative, and the second derivative term is applied to a down-sampled computation of the error derivative. This choice is motivated by the observation that the disturbances in the printing process mainly have two distinct origins: the faster machine motion system (characterized by a small time constant) and the slower filament extrusion system (having a larger time constant). Thus the standard (more responsive) derivative will account for the disturbances in the former system, while the down-sampled (more damped) derivative will account for the disturbances in the latter.
On a discrete and non-real-time system, this controller can be implemented iteratively as
\begin{equation}
v[k] = K_p e[k] + K_i e_i[k] + K_d e_d[k] + K_{dd} e_{dd}[k] \,,
\end{equation}
where
\begin{align}
e[k] &= F_r - F[k] \,, \\
e_i[k] &= e_i[k-1] + (t[k]-t[k-1]) e[k] \,,\\
e_d[k] &= \frac{e[k]-e[k-1]}{t[k]-t[k-1]} \,, \\
e_{dd}[k] &= \frac{e[k]-e[k-d]}{t[k]-t[k-d]} \,.
\end{align}
We have used $F_r$ to denote the reaction force reference, $F$ for the measured reaction force, and $t$ for the clock time. The control input $v$ corresponds to the velocity at which we actuate the filament driving wheel in the extruder. The notation $\cdot[k]$ indicates that a variable has been measured or computed during the last iteration of the controller; $\cdot[k-1]$ corresponds to the previous iteration, and $\cdot[k-d]$ to $d$ iterations in the past. $K_p$, $K_i$, $K_d$, and $K_{dd}$ are the controller tuning parameters that correspond to the proportional, integral, derivative, and down-sampled derivative parts respectively.

\subsection{Experimental setup} \label{sec:setup}

\begin{figure}[htbp]
\centerline{\includegraphics[width=0.49\textwidth]{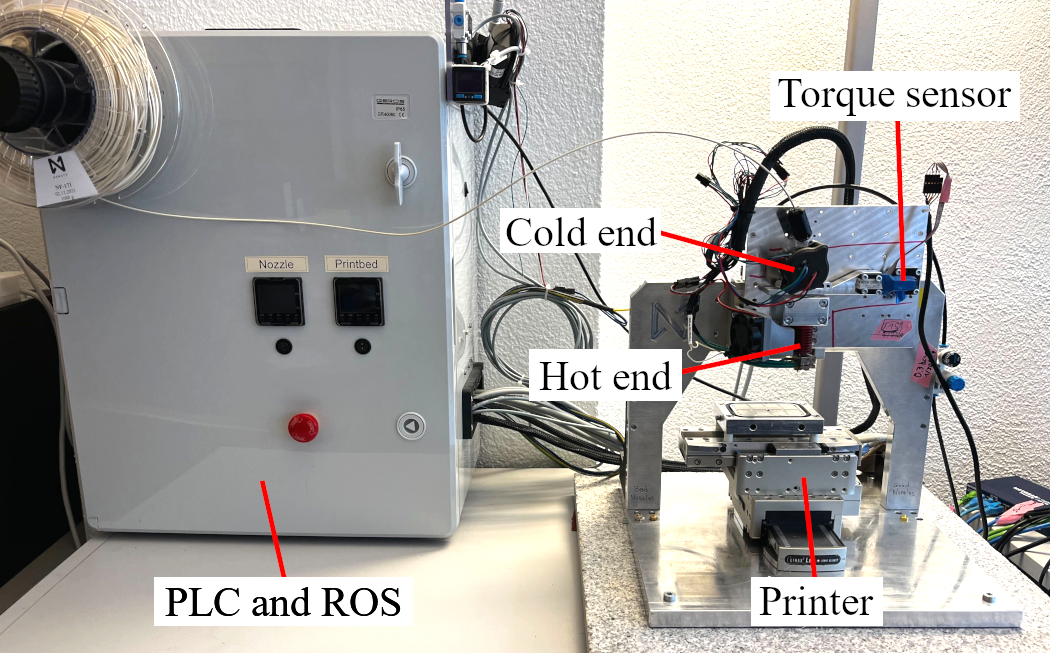}}
\caption{Custom-built FFF printer used for Force Controlled Printing}
\label{fig:printer}
\end{figure}

Prints were conducted on a custom-built Cartesian FFF machine actuated by Linax linear motors. The motors have an maximum speed of \SI{2}{m/s}, a maximum acceleration of \SI{40}{m/s^2} and an accuracy of \SI{1}{\micro\meter}. The LCP material was printed by using a Mellow Sherpa Micro
extruder (with LDO-20-8T stepper motor) customized as detailed in Sec.\ \ref{fig:printhead}. The complete hardware setup is shown in \cref{fig:printer}. During all experiments (unless differently specified) we utilized a nozzle diameter (and nominal line width) of \SI{0.15}{mm}, a nominal layer height of \SI{50}{\micro\meter}, a feed rate of \SI{100}{mm/s}, a nozzle temperature of \SI{300}{\celsius}, and a bed temperature of \SI{170}{\celsius}. The torque sensor utilized to measure the reaction force is a custom sensor produced by Bota Systems AG\footnote{\url{https://www.botasys.com/}}. The system is controlled by a programmable logic controller (PLC) using Beckhoff\footnote{\url{https://www.beckhoff.com/}} software connected to a computer running ROS2~\cite{ros2}. Details about the configuration can be found in \cite{guidetti2024print}.

All the part measurements and pictures were obtained with a Keyence LM-X100L digital microscope, set at a $200\times$ magnification. The microscope software was used to measure the width of individually printed lines by using edge detection.

\section{Calculation} \label{sec:calc}

\subsection{Controller Tuning and Validation}
We tune the extrusion controller to track a constant reaction force reference while rejecting disturbances. During a print, a well-tuned controller minimizes the accumulated difference between the measured reaction force $F$ and the reaction force reference $F_r$. Formally, we compute the controller performance during a print where $K$ total force measurements are recorded as the root-mean-square-error (RMSE)
\begin{equation}
J = \sqrt{\frac{\sum_{k=1}^K \left(F_r-F[k]\right)^2}{K}} \,.
\end{equation}
For a fully specified set of trajectories, machine feed rate, extrusion temperature, nozzle diameter, and reaction force reference, the optimal controller is the one minimizing $J$. In this work, we carefully tune the controller to achieve satisfyingly low values of $J$ during the general use of the printer (i.e. printing various shapes at a reaction force reference of \SI{0.2}{N}) with a fixed feed rate, extrusion temperature, and nozzle diameter. The tuning is conducted using the automated method in \cite{guidetti2024print}. The optimal controller parameters are listed in Table \ref{tab:control_parameters}.

\begin{table}[htbp]
\centering
\caption{Controller parameters used for prints conducted at a feed rate of \SI{100}{mm/s}, a temperature of \SI{300}{\celsius} and with a nozzle diameter of \SI{0.15}{mm}}
\renewcommand{\arraystretch}{1.3}
\begin{tabular}[h]{@{}c c c c c@{}}
\toprule
$K_{p}$ & $K_{i}$ & $K_{d}$ & $K_{dd}$ & $d$ \\
\midrule
44.72 & 7.22 & 2.25 & 1.12 &10 \\
\bottomrule
\end{tabular}
\label{tab:control_parameters}
\end{table}

\begin{figure}[htbp]
\centerline{\includegraphics{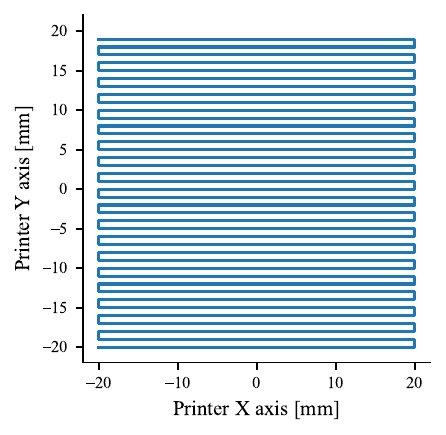}}
\caption{\emph{Snake} trajectory, used as a benchmark in this work. The trajectory is a continuous raster composed of 40 equally spaced \SI{40}{mm} long lines (oriented in the $X$ direction) connected by a \SI{1}{mm} long segment (in the $Y$ direction). Printing begins at $X=\SI{-20}{mm}, Y=\SI{-20}{mm}$, and finishes at $X=\SI{-20}{mm}, Y=\SI{19}{mm}$.}
\label{fig:snake}
\end{figure}

\begin{figure}[htbp]
\centerline{\includegraphics{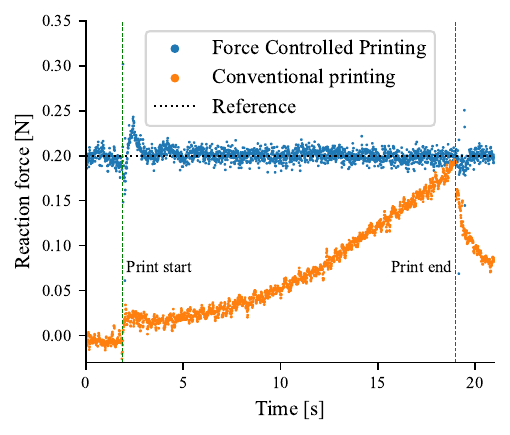}}
\caption{Comparison between the measured reaction forces during conventional printing and Force Controlled Printing. Both experiments were conducted on the same trajectory (shown in Fig.\ \ref{fig:snake}) at a feed rate of \SI{100}{mm/s}, a temperature of \SI{300}{\celsius}, a layer height of \SI{50}{\micro\meter}, and with a nozzle diameter of \SI{0.15}{mm}. Note that the force reference is not used by the open-loop print.}
\label{fig:control_validation}
\end{figure}

To validate the tuning and demonstrate the performance of the selected controller, we print the same trajectory twice, first using the conventional open-loop approach to extrusion and next with the tuned closed-loop controller. Details about the utilized hardware are given in \cref{sec:results}. The open-loop print acts as a benchmark showing the performance of a non-controlled conventional print. The printed trajectory is shown in \cref{fig:snake}, and the two approaches deposit beads of comparable width (approximately \SI{0.075}{mm}). We compare the reaction force measurements obtained during two experiments in \cref{fig:control_validation}. In the open-loop case, the material flow rate is constant throughout the print. The extruder motor rotates at a constant speed that was selected offline by a standard slicer. The reaction force measurement reflects the fact that the conventional open-loop approach produces long-lasting force transients: the extrusion force cannot reach a steady state in the time required to print the trajectory (the implications of this phenomenon are discussed in \cref{sec:disc}). In FCP, the tuned controller continuously adapts the extruder motor speed to maintain the reaction force close to the reference of \SI{0.20}{N}. It is possible to notice slight disturbances caused by the start and the end of the print, but the measured force generally tracks the reference closely. During the print, the closed-loop experiment has a $\text{RMSE}=\SI{9.14e-3}{N}$. This value amounts to \SI{4.6}{\percent} of the reference magnitude, and indicates that the controller is suitably tuned\footnote{The open-loop approach to extrusion is solely based on the theoretical calculation of the material flow rate required to produce a desired bead width, for a given bead height and printing feed rate. As the method is completely unrelated to following a reference reaction force, computing $J$ for the open-loop case is meaningless}. 

\section{Experiments and Results}
\label{sec:results}

To evaluate the performance of FCP, we have designed a set of experiments that highlight the different advantages of this approach. First, we demonstrate the existence of a strong correlation between the measured reaction force and the width of a deposited line. Then, we show how controlling the extrusion to maintain a constant reaction force during the print allows one to deposit lines of predefined width even at inconsistent layer heights. Finally, we show how the proposed method consistently produces high-quality parts under significant disturbances caused by filament slippage or incorrect print bed leveling.

\subsection{Line width to force correlation}

\begin{figure}[htbp]
\centerline{\includegraphics{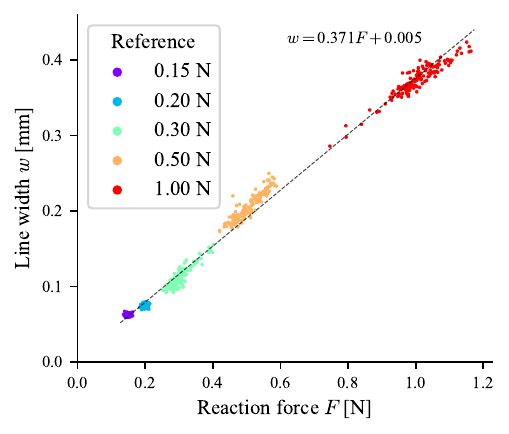}}
\caption{Measured line width and corresponding reaction force obtained in five separate experiments. In every experiment, a snake pattern was printed while tracking a different reaction force reference. Each experiment was analyzed with 140 uniformly spaced measurements.
All experiments were conducted at a feed rate of \SI{100}{mm/s}, a temperature of \SI{300}{\celsius}, a layer height of \SI{50}{\micro\meter}, and with a nozzle diameter of \SI{0.15}{mm}.
The linear fit has $R^2 = 0.99$ and $\text{RMSE} = \SI{9.24e-3}{mm}$.
}
\label{fig:scatter_force_width}
\end{figure}

The first study focuses on analyzing the correlation between the reaction force measured with our setup and the printed line width.
This is a key step in characterizing the geometrical properties of printed parts via in situ measurements, which enables the design of in-layer control architectures. 
We use FCP to print five separate snake patterns, each using a different reference reaction force. Specifically, we target references of \SI{0.15}{N}, \SI{0.20}{N}, \SI{0.30}{N}, \SI{0.50}{N}, and \SI{1.00}{N}. In all five experiments, we utilize the controller that has been tuned for the \SI{0.20}{N} case, whose performance is shown in \cref{fig:control_validation}, and print a snake pattern. The print bed is leveled and set to achieve a constant line height of \SI{0.05}{mm}. 
In conventional open-loop printing, the line height is given by the distance between the nozzle tip and the print bed, as the extruded material fills the void between the two elements. We use the same approach to set the layer height in our experiments and use FCP during prints.

For each of the five experiments, we measure the deposited line width at 140 uniformly spaced locations along the print trajectory. We then plot each width measurement in conjunction with the reaction force recorded while printing at the measurement location. The combined results from the five experiments are shown in \cref{fig:scatter_force_width}. 
Line width and extrusion reaction force show a strong linear correlation. We fit a linear regression to the data and evaluate the goodness of the linear fit. The metrics amount to $R^2 = 0.99$ and $\text{RMSE} = \SI{9.24e-3}{mm}$, which indicate an almost perfect linear relation and make line width predictable with high accuracy by using reaction force. 

In the experiments conducted with a force reference of \SI{0.30}{N}, \SI{0.50}{N}, and \SI{1.00}{N}, the performance of the controller deteriorates, and the measurements are distributed in a wider spread.
This reduction in performance is explained by the mismatch of tuning at \SI{0.20}{N} and testing at different references.
Nonetheless, the linear relation between reaction force and line width holds strongly. 
This indicates that with suitable controller tuning, such as the case for references of \SI{0.15}{N} and \SI{0.20}{N}, it is possible to extrude lines of any desired constant width simply by tracking the corresponding reaction force.
This result brings in a key insight into the material extrusion additive manufacturing literature and provides a new framework for process design as we demonstrate next.

\subsection{Variable layer height} \label{sec:var_height}

\begin{figure}[htbp]
\centerline{\includegraphics{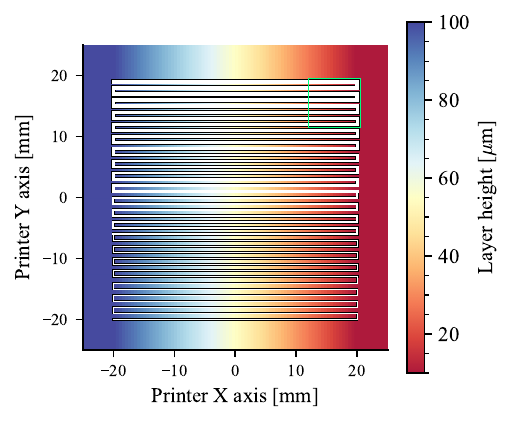}}
\caption{Design of the experiment for the variable layer height study. We print the \emph{snake} trajectory on a tilted print bed and use different force references. The bed tilting is indicated by the color gradient in the figure background. It produces a linearly variable layer height along the $X$ direction, ranging from a minimum of \SI{10}{\micro\meter} (at $X=\SI{20}{mm}$) to a maximum of \SI{100}{\micro\meter} (at $X=\SI{-20}{mm}$); the layer height of \SI{50}{\micro\meter} used in the previous studies is achieved close to $X=\SI{0}{mm}$. The variation in force reference is depicted with a change in the stroke width size of the white trajectory line. The print begins at $X=\SI{-20}{mm}, Y=\SI{-20}{mm}$ with a force reference $F_r = \SI{0.20}{N}$, and the reference is increased by \SI{0.10}{N} every seven raster passes, producing a stepwise increase in force throughout the print which terminates at $F_r = \SI{0.70}{N}$. The experiment was analyzed by measuring at 140 uniformly spaced locations. The green square indicates the area where the print was executed with the largest reaction force and smallest layer height.}
\label{fig:tilted_bed_case}
\end{figure}

In the next study, we analyze the performance of FCP when printing in sub-optimal conditions. 
In particular, we simulate the case where the print bed is not correctly leveled and thus the distance between the nozzle tip and the bed varies during the layer deposition. To make the task harder, we take no corrective action while printing, and just follow the same extrusion strategy we would use on a leveled bed. This is analogous to the case commonly encountered in conventional printing, in which the print bed is often warped or not parallel to the printer's horizontal motion plane, and this continuous change in layer height disturbs the printing process. 
Effects of spatial disturbances on the print input and geometry have been previously recorded~\cite{balta2021layer}.
Here, we demonstrate FCP as an effective way to compensate for such disturbances using run-time feedback.
We conduct this experiment with different force references in order to observe the joint effect of reaction force and variable layer height on the printed lines. \Cref{fig:tilted_bed_case} illustrates the experimental settings used in the study. 

\begin{figure}[htbp]
\centerline{\includegraphics{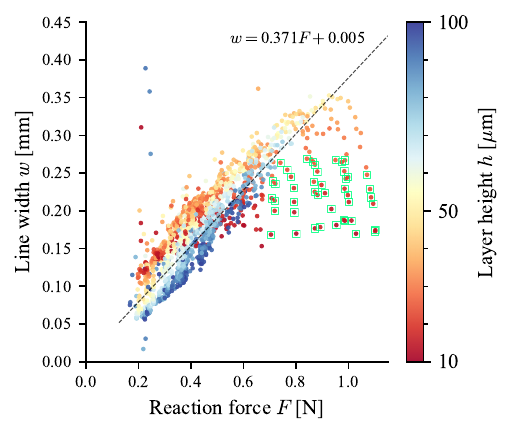}}
\caption{Results of the variable layer height study. Points coloring depends on the local layer height and corresponds to the color map used in Fig.\ \ref{fig:tilted_bed_case}. Points depicted in a green square belong to the region highlighted with a green square in Fig.\ \ref{fig:tilted_bed_case}. The parameters of the dashed line indicating the linear relation between reaction force and line width were identified using the data of Fig.\ \ref{fig:scatter_force_width} (i.e. printed at a nominal layer height of \SI{50}{\micro\meter}).}
\label{fig:tilted_bed_results}
\end{figure}

\begin{figure*}[htbp]
\centerline{\includegraphics{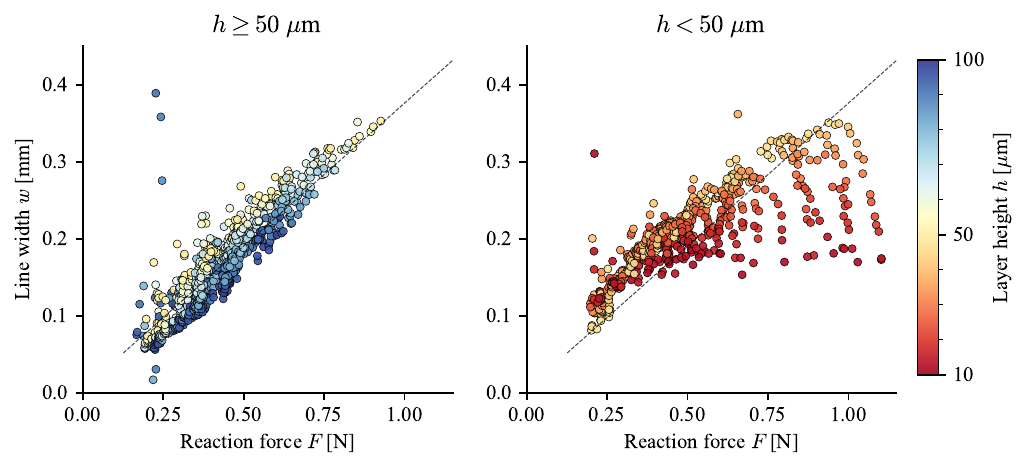}}
\caption{Results of the variable layer height study, separated based on the layer height. The left and right plots depict the measurements conducted in the region with a layer height larger than nominal and smaller than nominal, respectively.}
\label{fig:tilted_split}
\end{figure*}

The aggregated results of the experiment are illustrated in \cref{fig:tilted_bed_results}. The results show that the linear relation between the reaction force and the obtained line width is essentially not affected by the variation in layer height, as the points are well distributed along the linear fit that was obtained in \cref{fig:scatter_force_width}. One exception is produced in the region highlighted in green in \cref{fig:tilted_bed_case}, where the reference force is very large and the layer height very small. The points belonging to the highlighted region were also highlighted in \cref{fig:tilted_bed_results}, and they all fail to follow the previously identified linear relation between reaction force and line width. Specifically, we observe the limitations of the proposed approach: when the layer height is reduced excessively, the reaction force increase is not matched by a line width increase, as the latter saturates. In \cref{fig:tilted_split} it is possible to observe this phenomenon more clearly. Additionally, it is noticeable that printing at layer heights larger than the nominal height of \SI{50}{\micro\meter}, the force-width linear relation is excellent. For layer heights smaller than \SI{50}{\micro\meter}, the line width appears to be slightly larger than expected at low reaction forces and to saturate at high reaction forces. This phenomenon becomes more prevalent as the layer height becomes progressively smaller.

To produce a clear comparison between the conventional open-loop approach to extrusion and FCP, we have printed two \emph{snake} trajectories on a tilted bed using the two approaches. The bed tilting was set so that the layer height would vary linearly along the Y direction, from a maximum of \SI{100}{\micro\meter} (at $Y=\SI{-20}{mm}$, the beginning of the \emph{snake}) to a minimum of \SI{10}{\micro\meter} (at $Y=\SI{20}{mm}$, the end of the \emph{snake}). As previously discussed, in FFF improper leveling is not an easy-to-detect disturbance. Accordingly, no corrective action was taken during the experiments. In both cases, the print was conducted as it would on a flatbed. For FCP, a reaction force reference of \SI{0.55}{N} was set, to obtain a line width of \SI{0.2}{mm}. In the conventional open-loop approach, the extruder drive was set to rotate at constant speed. The speed was set to match the averaged rotational speed that the closed-loop approach would utilize when printing on a flat bed, at a layer height of \SI{50}{\micro\meter} and with a reaction force reference of \SI{0.55}{N}. This makes the two approaches extrude approximately the same amount of material in the ideal case where the bed is well leveled at \SI{50}{\micro\meter}. 

\begin{figure}[htbp]
\centerline{\includegraphics{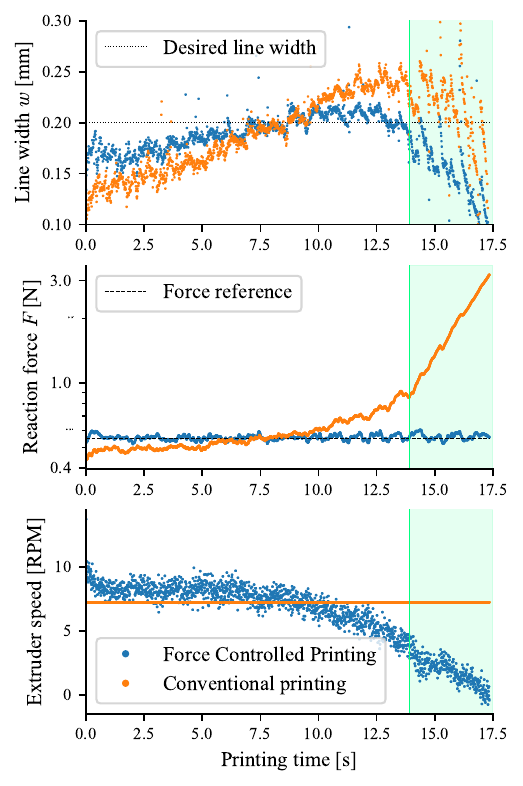}}
\caption{Comparison of line width, reaction force, and extruder speed for a conventional and a force-controlled print of the \emph{snake} on a tilted bed. The layer height is \SI{100}{\micro\meter} at the beginning of the print and decreases linearly (in the $Y$ direction of the \emph{snake} pattern) until \SI{10}{\micro\meter}. Measurements are plotted as a function of the printing time to follow the data evolution as the \emph{snake} print progresses. The green overlay indicates the region where the layer height is too small for the linear relation between force and line width to hold (as evaluated in Sec. \ref{sec:var_height}).}
\label{fig:open_vs_closed_tilt}
\end{figure}

\begin{figure*}[htbp]
\centerline{\includegraphics{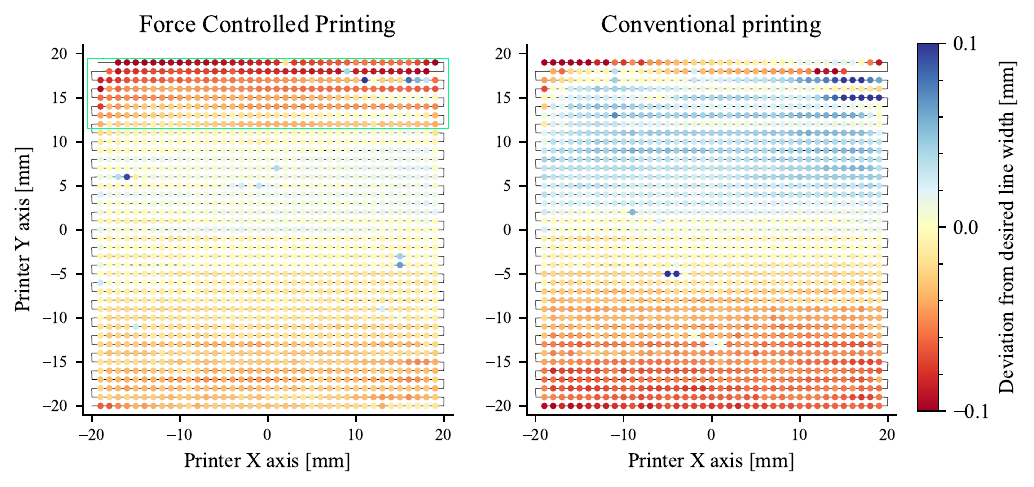}}
\caption{Comparison of line width for a conventional and a force-controlled print of the \emph{snake} on a tilted bed. The layer height is \SI{100}{\micro\meter} at the $Y=\SI{-20}{mm}$ and decreases linearly (in the $Y$ direction) until \SI{10}{\micro\meter} at $Y=\SI{20}{mm}$. The desired line width is $w=\SI{0.2}{mm}$. The green overlay indicates the region where the layer height is too small for the linear relation between force and line width to hold (as evaluated in Sec. \ref{sec:var_height})}
\label{fig:open_vs_closed_tilt_bis}
\end{figure*}

The results are shown in \cref{fig:open_vs_closed_tilt,fig:open_vs_closed_tilt_bis}, where we can see the different behavior of the two approaches when printing on an uneven bed. The print begins with a very large layer height: the closed-loop approach extrudes faster to maintain the reaction force at the reference, while in open-loop the prescribed extrusion speed produces a lower force. In the middle of the print, where the layer height is close to \SI{50}{\micro\meter}, both approaches perform similarly as there is little to no disturbance induced by the bed. In the final part of the print, where the layer height is very small, the open-loop approach produces an extremely large reaction force, and the extruder wheel was observed to slip on the filament; the closed-loop approach reduces the extruder speed progressively to regulate the reaction force. Confirming the previously established correlation between reaction force and line width, FCP produces a more constant line width, as desired. 

The disturbances induced by the tilted bed are well rejected with FCP. However, it is possible to observe that very low layer heights (which are achieved at the very end of the print, from ca. \SI{14}{s} onward) invalidate the linear relation between force and line width, as already discussed in~\cref{sec:var_height}. This effect worsens the printing performance of FCP, which fails to produce the desired line width in this regime. Despite this, using the desired line width of \SI{0.2}{mm} to compute the RMSE, we see that the open-loop RMSE of \SI{0.042}{mm} is reduced to \SI{0.031}{mm} when utilizing FCP. After removing the measurements obtained in the region where the performance of the closed-loop approach is compromised, we obtain an RMSE of \SI{0.041}{mm} for conventional printing and of \SI{0.019}{mm} for FCP, which corresponds to a \SI{54}{\percent} decrease in RMSE. These results are summarized in \cref{tab:open_vs_closed_tilt}.

\begin{table}[htbp]
\centering
\caption{Comparison of line width RMSE when printing on a tilted bed with Force Controlled Printing and with conventional open-loop printing. The results reported as \emph{linear region} are computed after removing the data collected where the layer height is too small for the linear relation between force and line width to hold. The removed data belongs to the green overlay of \cref{fig:open_vs_closed_tilt}.}
\renewcommand{\arraystretch}{1.3}
\begin{tabular}[h]{@{}l r r@{}}
\toprule
& FCP & Conventional \\
\midrule
\makecell[l]{RMSE [\SI{}{mm}]\\ (entire experiment)} & \bf 0.031 & 0.042 \\
\makecell[l]{RMSE [\SI{}{mm}]\\ (linear region)} & \bf 0.019 & 0.041 \\
\bottomrule
\end{tabular}
\label{tab:open_vs_closed_tilt}
\end{table}

\subsection{Part Print}
In this final set of experiments, we are interested in analyzing the performance of FCP when manufacturing a full solid 3D part. 
The geometry we utilize is a right prism whose base can be seen in \cref{subfig:slippage_CL} and named \emph{star}. 
The geometry includes regular rectangular features and irregular shapes and corners with variable print line lengths for the infill. Our choice was to present a challenging print geometry that would be representative of practical geometries while feasible to print and analyze in our experimental setup. 
The printing trajectories are generated to produce a parallel infill with lines spaced \SI{0.15}{mm}, and the printed part is composed of eight identically repeated layers. As done previously, prints are executed using the standard settings reported in Sec. \ref{sec:setup}.

The performance of FCP is compared with conventional open-loop printing in two different scenarios. First, we artificially induce slippage between the filament driving wheel and the filament. Second, we print on a poorly leveled bed. The open-loop conventional prints utilize the g-code file generated by a state-of-the-art commercial slicer~\cite{prusaslicer}. Thus, the extruder commands are fully synchronized with the machine motion as normally done in conventional FFF.

\subsubsection{Driving wheel slippage}

The filament driving wheel is normally held firmly in contact with the filament by a spring-loaded screw. If the screw is not correctly tightened, or if it becomes loose while operating the machine, there is slippage between the driving wheel and the filament. 
The slippage is more pronounced when printing at high feed rates. 
When slippage occurs, a conventional open-loop extruder deposits a reduced amount of material than planned, producing defects in the printed part \cite{baechle2022failures, greeff2017closed}. 

\begin{figure*}[htbp]
\centering
    \begin{subfigure}[t]{0.49\textwidth}
         \centering
         \includegraphics[scale = 0.9]{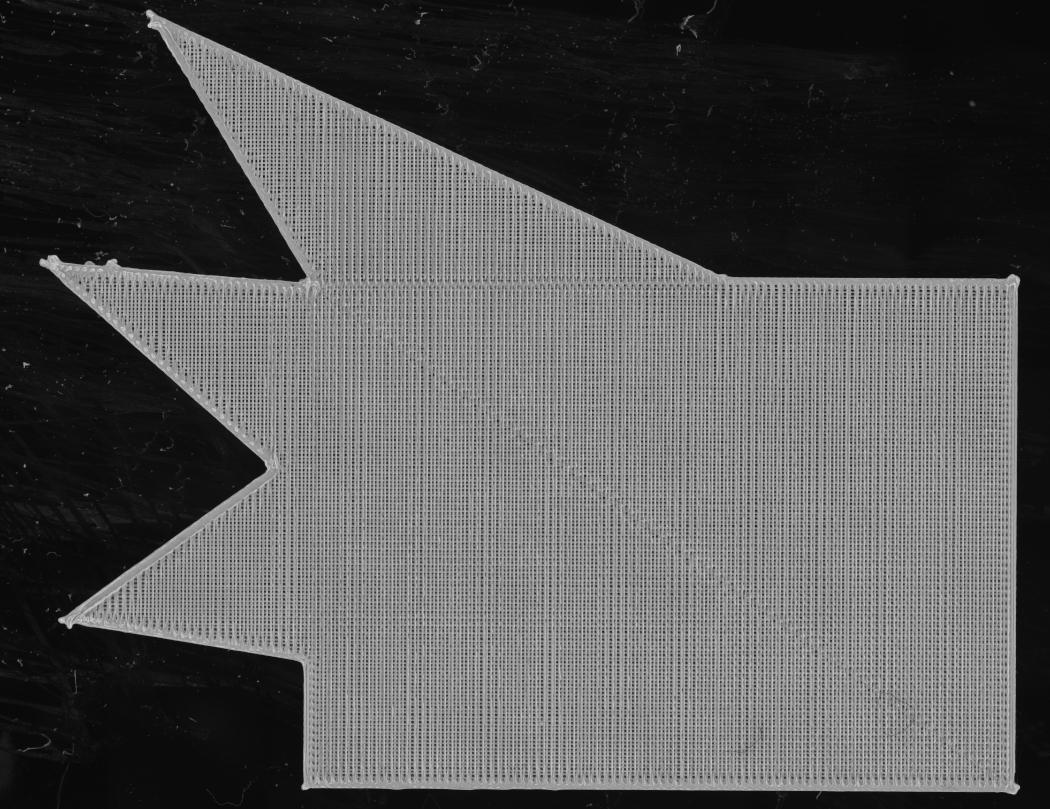}
         \caption{Force Controlled Printing}
         \label{subfig:slippage_CL}
     \end{subfigure}
     \hfill
     \begin{subfigure}[t]{0.49\textwidth}
         \centering
         \includegraphics[scale = 0.9]{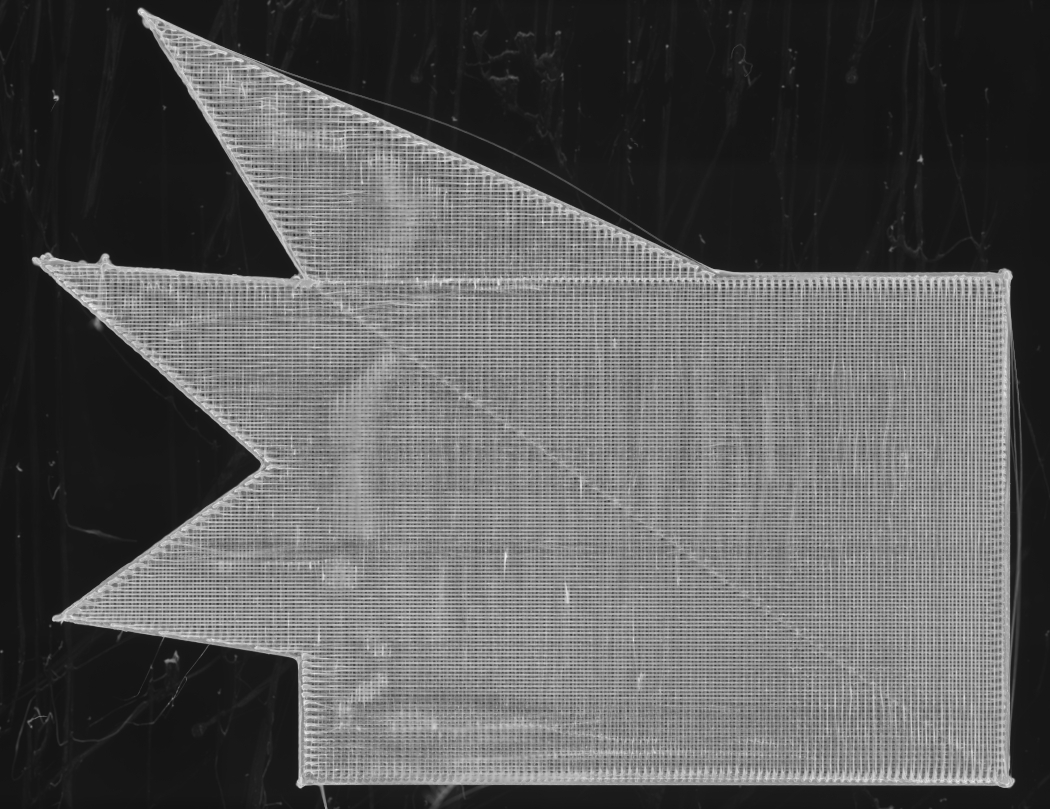}
         \caption{Conventional printing}
         \label{subfig:slippage_OL}
     \end{subfigure}   
\caption{Comparison of \emph{star} prints with filament driving wheel slippage}
\label{fig:slippage}
\end{figure*}

In \cref{fig:slippage}, we compare two prints of the \emph{star} produced after intentionally loosening the screw that holds the filament and driving wheel together. The extrusion is set both in the closed-loop controller and in the open-loop slicer to produce beads of width \SI{0.075}{mm}. The results show clearly how the part produced with open-loop printing (\cref{subfig:slippage_OL}) contains numerous slippage-related defects: poor inter-layer adhesion visible as lighter-colored halos in the center of the part, stringing around the perimeter on top and right edges, and over-extrusion at the intersection of infill and contour lines on the bottom-left and left edges. None of these defects are visible in \cref{subfig:slippage_CL}, showing that the closed-loop print strategy successfully uses reaction force feedback to increase the driving wheel speed and maintain the material flow constant even under severe slippage.

This result provides a practical example of how FCP can improve the current state of practice by enabling fault-tolerant control of FFF processes. As slippage causes underextrusion in most cases, research and practice have focused on detecting such regimes to correct them with heuristic approaches. Our framework provides a structured methodology to efficiently address such disturbances to the process by feedback control.

\subsubsection{Tilted print bed}

\begin{figure*}[htbp]
\centering
    \begin{subfigure}[b]{0.49\textwidth}
         \centering
         \includegraphics[height=2.1in]{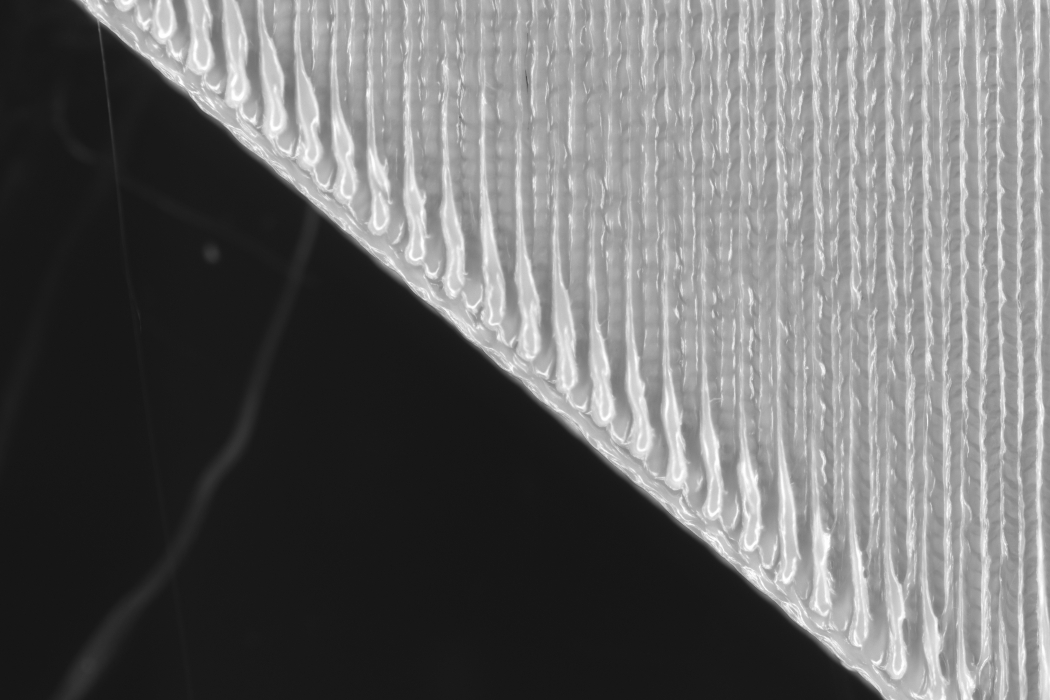}
         \caption{Force Controlled Printing}
         \label{subfig:convPLA}
     \end{subfigure}
     \hfill
     \begin{subfigure}[b]{0.49\textwidth}
         \centering
         \includegraphics[height=2.1in]{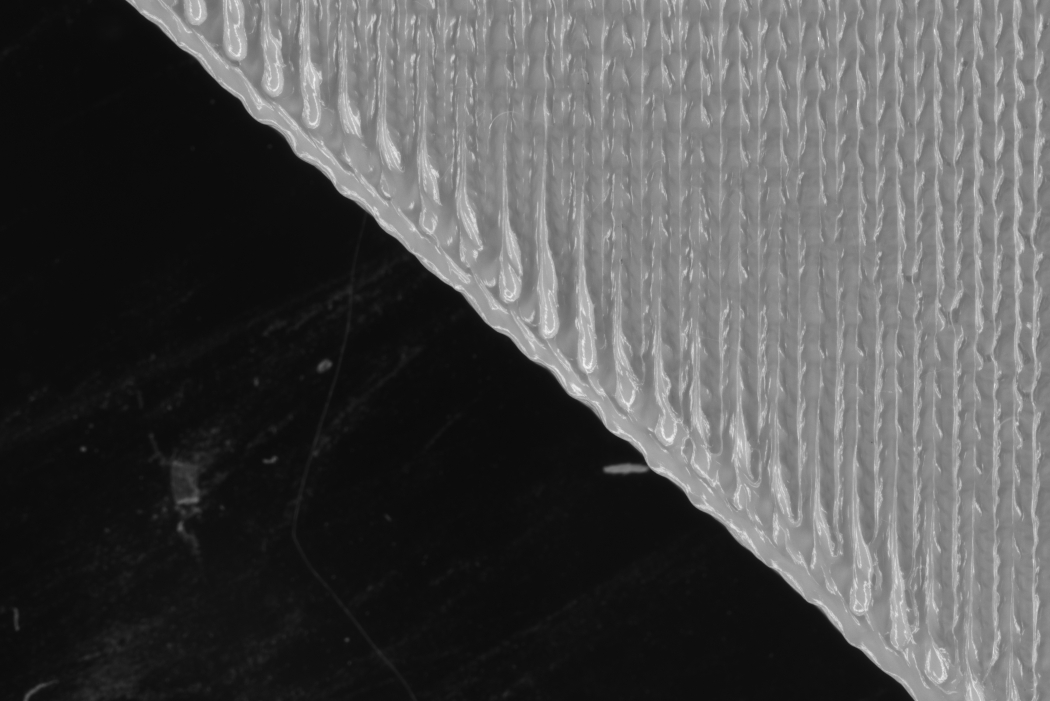}
         \caption{Conventional printing}
         \label{subfig:optPLA}
     \end{subfigure}
\caption{Comparison of contour line accuracy in the \emph{star}}
\label{fig:orth_comp}
\end{figure*}

\begin{figure*}[htbp]
\centering
     \begin{subfigure}[b]{0.49\textwidth}
         \centering
         \includegraphics[height=2.1in]{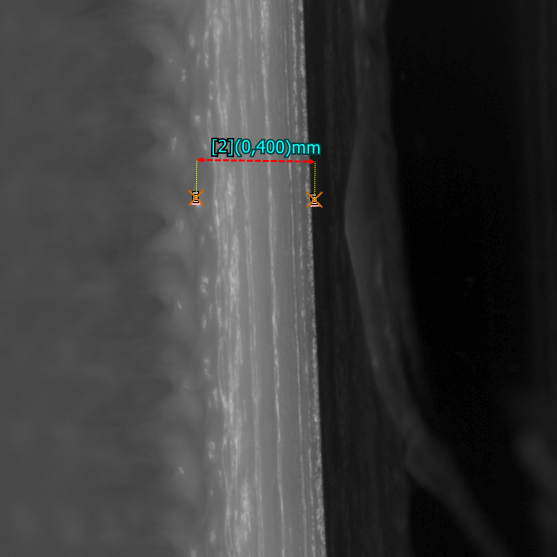}
         \caption{Side wall with FCP}
         \label{subfig:sidewall2_cl}
     \end{subfigure}
     \hfill
     \begin{subfigure}[b]{0.49\textwidth}
         \centering
         \includegraphics[height=2.1in]{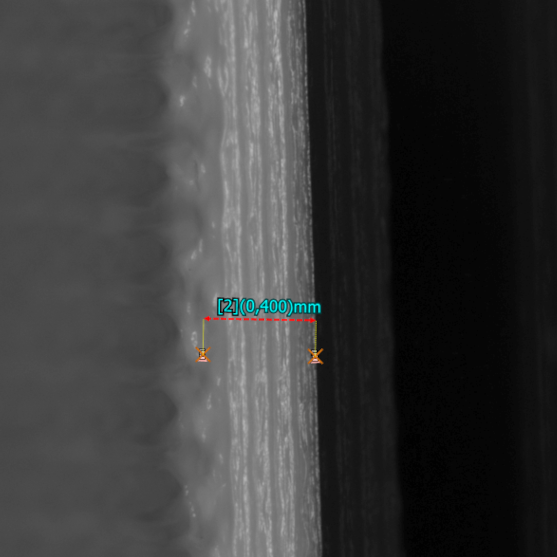}
         \caption{Conventional side wall}
         \label{subfig:sidewall2_ol}
     \end{subfigure}
         \vskip\baselineskip
     \begin{subfigure}[b]{0.49\textwidth}
         \centering
         \includegraphics[height=2.1in]{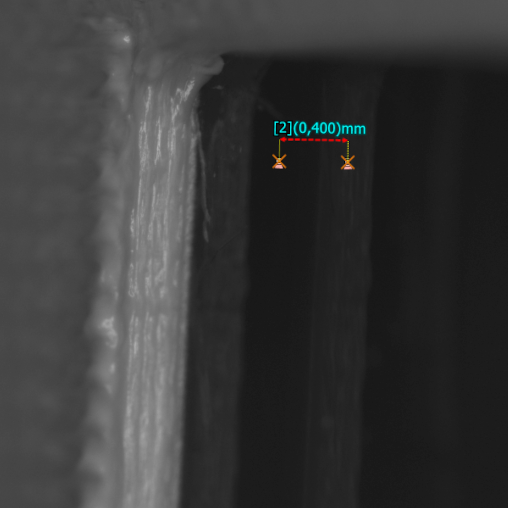}
         \caption{Side wall with FCP}
         \label{subfig:sidewall1_cl}
     \end{subfigure}
     \hfill
     \begin{subfigure}[b]{0.49\textwidth}
         \centering
         \includegraphics[height=2.1in]{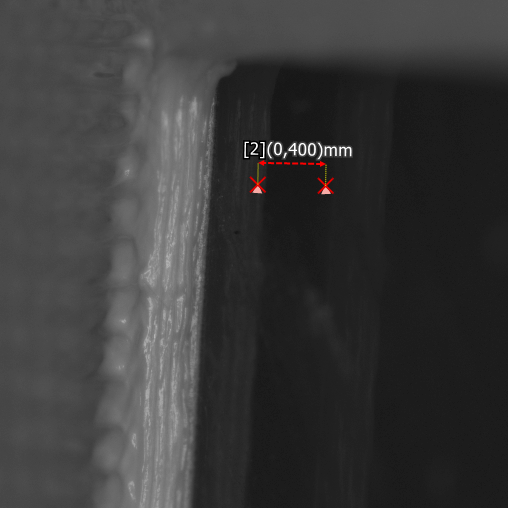}
         \caption{Conventional side wall}
         \label{subfig:sidewall1_ol}
     \end{subfigure}
     \vskip\baselineskip
     \begin{subfigure}[b]{0.49\textwidth}
         \centering
         \includegraphics[height=2.1in]{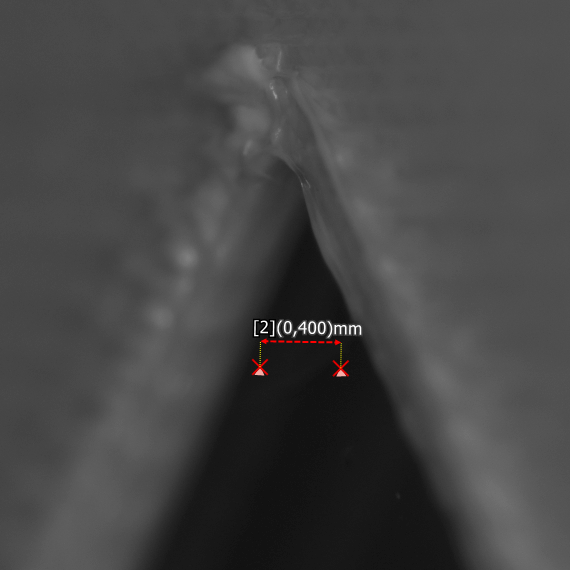}
         \caption{Contour line with FCP}
         \label{subfig:contour_cl}
     \end{subfigure}
     \hfill
     \begin{subfigure}[b]{0.49\textwidth}
         \centering
         \includegraphics[height=2.1in]{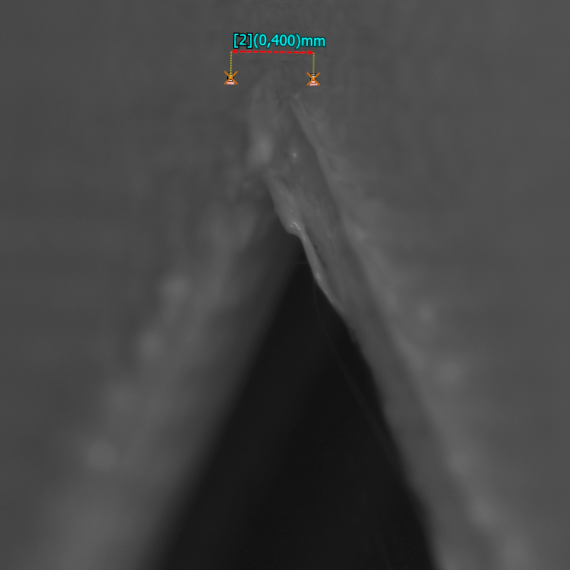}
         \caption{Conventional contour line}
         \label{subfig:contour_ol}
     \end{subfigure}
\caption{Comparison of part details in the \emph{star}. 
}
\label{fig:side_comp}
\end{figure*}

In this comparison between conventional printing and FCP, we print the star on a bed tilted as in \cref{fig:tilted_bed_case}, where the layer height changes linearly from \SI{100}{\micro\meter} to \SI{10}{\micro\meter} along the $X$ direction. Both the closed-loop extrusion controller and the open-loop slicer are set to produce lines of width \SI{0.15}{mm} (corresponding to \SI{100}{\percent} infill). The slicer computes the extrusion commands based on the nominal layer height of \SI{50}{\micro\meter}. The results are shown in \cref{fig:orth_comp,fig:side_comp}. 

\Cref{fig:orth_comp} shows a magnified view of the top surface of the star, where the precision of the part outline can be compared. FCP produces a very regular and straight outline, and the final part has a high geometric accuracy. With conventional extrusion, on the other hand, the outline of the part appears to be wavering, producing poor surface properties in the final part. This is due to the fact that the open-loop approach of conventional extrusion is unable to compensate for any variation in the printing conditions and that even a minor mismatch in the parameters used for extrusion planning produces over- or under-extrusion.

In \cref{fig:side_comp} we show selected features of the star with strong magnification. In particular, we observe the part side walls and the material residue left during contour deposition. The closed-loop approach adapts the extrusion process to the characteristics of the substrate upon which the polymer is deposited. On the other hand, when utilizing conventional printing, any imperfection in lower layers propagates to higher layers. This results in FCP producing smoother side walls (see \cref{subfig:sidewall2_cl,subfig:sidewall1_cl}) when compared to conventional printing. The contour lines of the top layer (see \cref{subfig:contour_cl}) appear to be of better quality than in the part produced with open-loop printing, where over-extrusion is visible (see \cref{subfig:contour_ol}).
 
\section{Discussion} \label{sec:disc}

We have shown in \cref{sec:results} numerous advantages produced by utilizing FCP instead of conventional open-loop extrusion. Our approach is able to
\begin{enumerate}
    \item Deposit lines with a desired width (over a large line width range) by selecting the corresponding reaction force reference;
    \item Maintain the correct line width while printing on a non-leveled surface;
    \item Produce high-quality prints even under significant disturbances to the process such as driving wheel slippage; and
    \item Produce parts with superior surface properties and geometrical accuracy.
\end{enumerate}
In practice, while high-quality printing is possible with a conventional open-loop approach, this comes at the cost of time-consuming parameters and machine tuning. The resulting performance is strongly affected by any minor error in the tuning process and by parametric drift in the equipment. In contrast, FCP produces high-quality parts without requiring hyper-precise configuration of the hardware and in a large range of conditions. 

Another significant benefit of FCP is the deposition efficiency that it can reliably achieve. In conventional printing, the nozzle diameter is used as nominal line width and achievable line width commonly ranges from \SI{75}{\percent} up to \SI{125}{\percent} of the nozzle diameter \cite{kechagias2022key}. We have shown in the experiments (see \cref{fig:tilted_bed_results}) that it is possible to utilize a \SI{0.15}{mm} nozzle to deposit lines of predefined width ranging from \SI{33}{\percent} up to \SI{233}{\percent} of the nozzle diameter, significantly extending the range of line widths achievable by a single nozzle. This marks a significant improvement over the existing approaches, which could translate to significantly faster print times in production.
Additionally, the material flow in the nozzle is at a steady state since the inception of the print (see \cref{fig:control_validation}), eliminating the need for material purging and the start-of-deposition defects which are typical in conventional open-loop extrusion. 

All the presented results have been achieved while printing with extremely large disturbances in the bed leveling. These modify the measured extrusion reaction force $F$ by acting on the $F_s$ component (see Eq.\ \eqref{eq:extr_force},\eqref{eq:balance}). By regulating the extrusion feed rate to track a constant force reference, more material is extruded when the substrate is far from the nozzle tip and vice versa. As a consequence, we can successfully produce the desired line width as well as high surface quality parts while varying the layer height between \SI{20}{\percent} and \SI{200}{\percent} of the nominal layer height. Such disturbances have been shown to strongly impair the performance of conventional open-loop extrusion, both in terms of line width accuracy and part surface quality~\cite{balta2021layer, percoco2021analytical, fischer2022improving}. In conventional printing, practitioners attempt to circumvent leveling issues by performing time-consuming bed leveling routines and using distance or contact sensors to create a height map of the print bed for compensation. With FCP, large leveling disturbances are dealt with automatically.

Our studies also highlighted the limitations of the proposed approach. We found that the linear relation between reaction force and line width, upon which FCP is based, no longer holds when printing at very wide lines at very small layer heights. While the physical causes of this phenomenon remain to be investigated, this behavior is to be expected. There is a maximum amount of material that can be forced out of the nozzle and into a very small gap for a given printing speed; exceeding this amount simply increases the measured reaction force without producing an increase in line width. In practice, we have observed this effect to happen only when printing simultaneously below \SI{50}{\percent} of the nominal layer height and above \SI{150}{\percent} of the nominal line width. These restrictions do not depend on our approach, rather, they are unavoidable consequences of the physical behavior of the extrusion process. Understanding such limits for a given process setup could improve process design in future work.

We believe it is important to highlight that, while the main motivation behind our approach lies in controlling the process, the proposed sensor setup can be used for process monitoring and part certification. By continuously measuring the reaction force during printing, it is possible to reconstruct the properties of the deposited lines at every location in the part. Note that this holds irrespective of the extrusion strategy, and thus can also be applied to conventional open-loop printing. Conventional FFF suffers, similar to many other AM processes, from a lack of clear and widely accepted part certification strategies. It is commonplace to perform destructive tests on a critical AM component and assume that an exact copy of the component will perform similarly. The limitation of this technique resides in the open-loop nature of AM processes, which limits repeatability: it is generally hard to produce a lot of parts with identical mechanical properties. We claim that the use of our reaction force sensor can advance the field of FFF part validation, which can be achieved by verifying that collected printing data show no anomaly.

\section{Conclusions}

We have introduced and studied the first comprehensive framework for Force Controlled Printing in FFF. In particular, we describe a custom-built extruder including a sensor that can measure the extrusion force in real-time. We utilize this measurement as feedback in a closed-loop control algorithm which continuously regulates the speed of the filament driving wheel to produce a desired force. We demonstrate how the extrusion force measured during the printing process is linearly correlated with the width of the deposited plastic beads. We utilize this finding to print lines of desired width, ranging from \SI{33}{\percent} up to \SI{233}{\percent} of the nozzle diameter. We show how when using Force Controlled Printing, the deposited line width remains largely unaffected by disturbances in the bed leveling comprised in a range of \SI{20}{\percent} to \SI{200}{\percent} of the nominal layer height. We compare Force Controlled Printing with conventional extrusion, where the commands to the extruder are pre-computed by a \emph{slicing} software and used in a feed-forward fashion. We demonstrate how Force Controlled Printing produces parts with superior surface properties and geometrical accuracy in the presence of process disturbances such as slippage or incorrect bed leveling. Our method also requires no modeling of the extrusion process and thus is not sensitive to hardware and parameter tuning errors that affect conventional extrusion.

\section*{Funding}

This work was supported by the Swiss Innovation Agency (Innosuisse, grant \textnumero 102.617) and by the Swiss National Science Foundation under NCCR Automation (grant \textnumero 180545).

\section*{Acknowledgments}

We acknowledge the support of NematX AG which provided the LCP material and experimental setup for this work. We also acknowledge the support of Bota Systems AG which provided the utilized torque sensor.



 \bibliographystyle{elsarticle-num} 
 \bibliography{cas-refs}





\end{document}